\documentclass[10pt,conference]{IEEEtran}

\usepackage{graphicx}
\usepackage{url}
\usepackage{booktabs}
\usepackage{array}
\usepackage{multirow}
\usepackage{listings}
\usepackage{xcolor}
\usepackage{caption}
\usepackage{subcaption}
\usepackage{balance}
\usepackage[hidelinks]{hyperref}

\lstdefinestyle{terminal}{
  basicstyle=\ttfamily\footnotesize,
  breaklines=true,
  frame=single,
  backgroundcolor=\color{gray!6},
  columns=fullflexible
}

\title{Security and Human-Centered Assessment of BACnet-Controlled DALI Infrastructure in an Educational Building Automation Testbed}

\author{
\IEEEauthorblockN{Ariton Verush}
\IEEEauthorblockA{
MSc in Computer Science\\
University of Bern\\
Bern, Switzerland\\
Email: ariton.verush@students.unibe.ch}
}

\begin{document}
\maketitle

\begin{abstract}
Building automation and control systems integrate heating, ventilation, air conditioning, lighting, sensing, and management functions through specialized communication protocols. While this integration enables flexible building operation, it also creates complex cyber-physical environments that are difficult to inspect, secure, and explain to new analysts. This paper presents a practical security and human-centered case study of a BACnet/IP building automation testbed with DALI lighting infrastructure, investigated during a domotics-oriented cybersecurity hackathon in Thun, Switzerland in April 2026. The study combines network-oriented enumeration, object-level inspection, physical rack analysis, and reflective HCI analysis of tool-supported learning. Using Yabe and BACteria, the work documents observable BACnet services, reconstructs structured object hierarchies, identifies room-level lighting-control paths, and maps BACnet objects to DALI group-level infrastructure. The analysis emphasizes that BACS assessment is not only a technical protocol task: it also requires usable tool interfaces, physical observability, interpretable naming conventions, and safe mental models for command priorities. The paper contributes a compact case study of BACnet/DALI exploration in an educational testbed and discusses implications for cybersecurity education, human-centered security tooling, and responsible experimentation in cyber-physical building environments.
\end{abstract}

\begin{IEEEkeywords}
BACnet, DALI, building automation, cybersecurity, HCI, cyber-physical systems, security education, Yabe, BACteria.
\end{IEEEkeywords}

\section{Introduction}

Building automation and control systems (BACS) are increasingly used to coordinate heating, ventilation, air conditioning, lighting, sensing, access control, and management services. BACnet is a standardized building automation communication protocol maintained under ASHRAE SSPC 135 \cite{bacnet_committee,ashrae_bacnet}. DALI, the Digital Addressable Lighting Interface, provides a standardized communication system for lighting-control devices \cite{dali_alliance}. Together, such technologies form cyber-physical infrastructures in which digital objects, network services, automation logic, and physical components are tightly coupled.

This coupling creates two simultaneous challenges. First, from a cybersecurity and networking perspective, exposed building automation services may reveal device identities, supported services, object structures, and control semantics. Prior surveys and cyber-physical reviews of BACS security emphasize that building automation environments have become important targets because they manage operationally relevant functions such as HVAC, lighting, and facility services \cite{graveto2022bacs,morales2024bas,li2023critical}. Second, from a human-computer interaction (HCI) perspective, these environments are difficult to understand: tools expose large object trees, protocol identifiers may not match physical labels, and visible physical effects may depend on automation priorities and system state.

The Cyber-Defence Campus BACS Hackathon described by armasuisse illustrates the growing role of realistic test environments for BACS security learning and experimentation \cite{armasuisse_bacs_2025}. Such environments allow participants to interact with automation controllers, field devices, sensors, lighting infrastructure, and management systems in constrained educational settings. This paper reports a focused case study from a domotics-oriented hackathon environment in Thun, Switzerland in April 2026. The analysis centers on a Siemens-based BACnet/IP rack infrastructure with DALI-connected lighting devices.

The contribution is deliberately half technical and half human-centered. Technically, the paper documents BACnet service enumeration, object-type visibility, structured-view traversal, and DALI group-level mapping. From an HCI perspective, the paper examines how tool design, naming conventions, rack labels, and physical feedback shape a participant's ability to reason safely about BACS infrastructure.

\section{Research Questions}

The case study is guided by three research questions:

\begin{itemize}
    \item \textbf{RQ1:} How can BACnet object hierarchies be reconstructed to identify physical control pathways in a building automation testbed?
    \item \textbf{RQ2:} What cybersecurity-relevant information becomes visible through BACnet enumeration and object-level inspection?
    \item \textbf{RQ3:} What human-centered challenges arise when novice or interdisciplinary participants map protocol-level objects to physical building infrastructure?
\end{itemize}

These questions avoid treating the exercise as a simple attack demonstration. Instead, the focus is on structured analysis, interpretability, and responsible learning in a realistic cyber-physical environment.

\section{Related Work}

\subsection{Building Automation Security}

BACS security has received increasing attention because modern buildings combine operational technology, IP-based networking, facility-management requirements, and cyber-physical control processes. Graveto et al. survey threats, attacks, and open issues in building automation and control systems, highlighting that these environments combine cyber and physical risk \cite{graveto2022bacs}. Morales-Gonzalez et al. review security concerns across building automation systems and discuss protocol-level weaknesses, attack surfaces, and defensive considerations relevant to technologies such as BACnet, KNX, Modbus, ZigBee, and Z-Wave \cite{morales2024bas}. Li et al. provide a cyber-physical review of building automation security, emphasizing threats across management, automation, and field layers, which is particularly relevant for systems that combine networked controllers with sensors, actuators, and room-level devices \cite{li2023critical}. Gasser et al. investigated publicly reachable building automation systems and discussed the security implications of exposed BACnet deployments \cite{gasser2017bacnet}. Together, these works show that BACS security must be analyzed not only at the network-service level, but also through object models, automation logic, protocol semantics, and physical device interactions.

\subsection{BACnet, BACnet/SC, and Secure Operation}

BACnet supports interoperability through a common object model and standardized services \cite{ashrae_bacnet}. However, BACnet/IP deployments require careful network design, segmentation, and service exposure management. BACnet Secure Connect (BACnet/SC) was introduced to add secure, encrypted communication suitable for modern IP infrastructures \cite{bacnet_sc_whitepaper}. Although the present case study focuses on BACnet/IP exploration, BACnet/SC remains important as a broader mitigation direction for modern BACS.

\subsection{Human-Centered Cybersecurity and Cyber Ranges}

Human-centered cybersecurity argues that security effectiveness depends not only on technical controls but also on user understanding, usability, and interaction context \cite{grobler2021human}. Cyber ranges and security testbeds are increasingly used to support hands-on learning, controlled experimentation, and structured cybersecurity training scenarios \cite{yamin2020cyberranges}. BACS testbeds extend this idea into cyber-physical domains, where learners must connect digital evidence to physical devices, environmental behavior, and operational constraints.

\section{Experimental Context and Methodology}

\subsection{Hackathon Context}

The case study was conducted during the Domotics Hackathon April 2026 in Thun, Switzerland, held by the Cyber-Defence Campus (CYD). Publicly available documentation from armasuisse describes a related CYD Campus BACS Hackathon in Thun, focused on building automation cybersecurity and involving vendors, automation specialists, academics, and cybersecurity participants \cite{armasuisse_bacs_2025}. The present paper uses that public page only as background context for the BACS hackathon setting. The device observations, screenshots, and object mappings reported here are based on the author's own participation records from the April 2026 event.

\subsection{Tools}

Two tools were used:

\begin{itemize}
    \item \textbf{Yabe}: a graphical BACnet explorer used for manual browsing, object-tree inspection, property validation, and visual correlation of object names with device behavior.
    \item \textbf{BACteria}: a command-line BACnet testing and enumeration framework used here primarily for device fingerprinting, service discovery, object inspection, and structured-view traversal.
\end{itemize}

The workflow emphasized observation, mapping, and documentation. Control-relevant interactions were interpreted through BACnet priority semantics and remained within the authorized educational testbed.

\subsection{Evidence Sources}

The paper uses screenshots, rack diagrams, rack photographs, command-line output, and participant notes. The relevant evidence includes: full testbed diagrams, Rack 2 architecture, a Rack 2 device inventory, front and rear rack photographs, BACteria command output, Yabe inspection screenshots, and records of BACnet object traversal.

\section{Experimental Environment}

\begin{figure*}[t]
\centering
\includegraphics[width=0.95\textwidth]{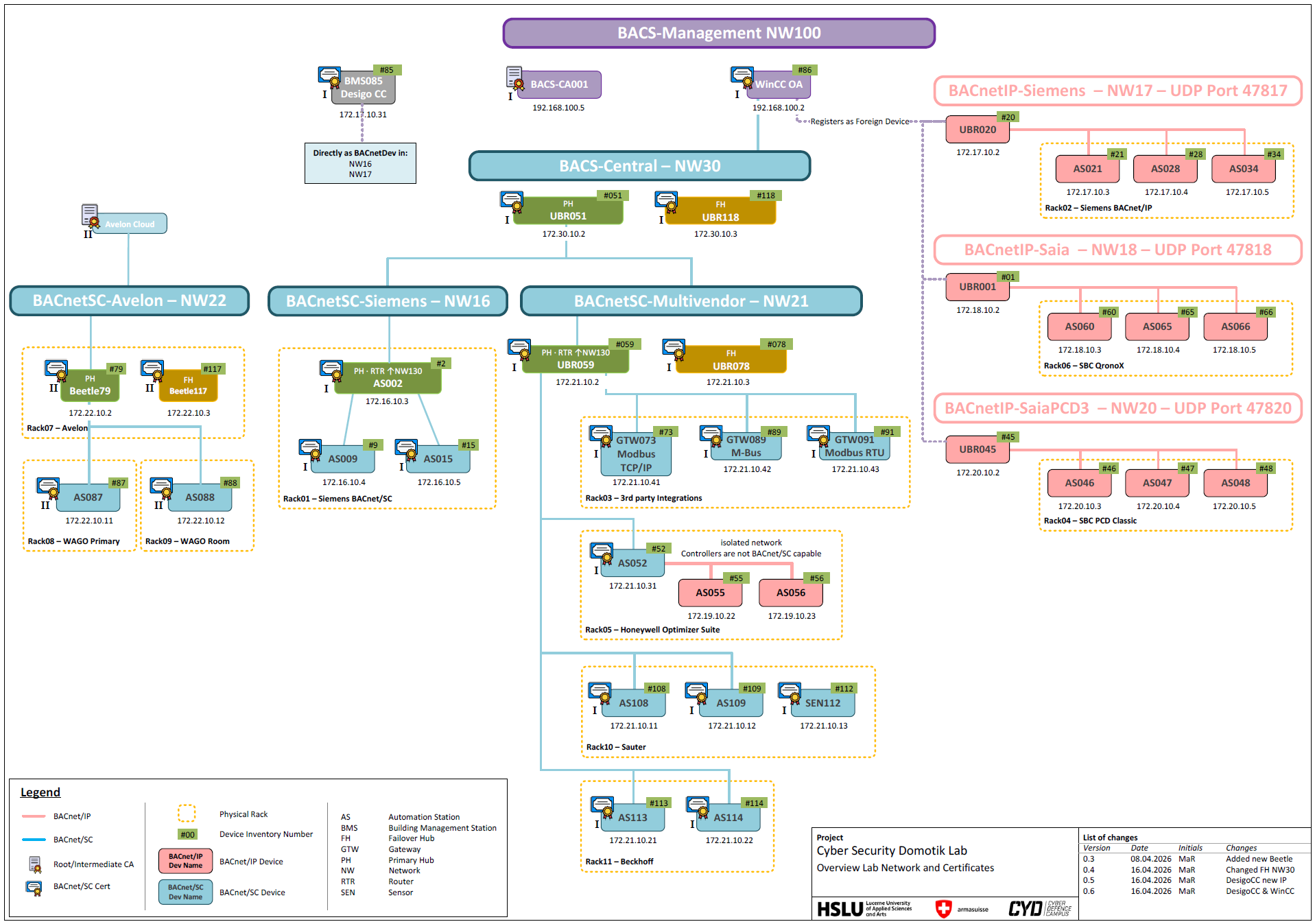}
\caption{Overview of the building automation testbed and network segmentation used during the domotics hackathon.}
\label{fig:overview}
\end{figure*}

\begin{figure*}[t]
\centering
\includegraphics[width=0.95\textwidth]{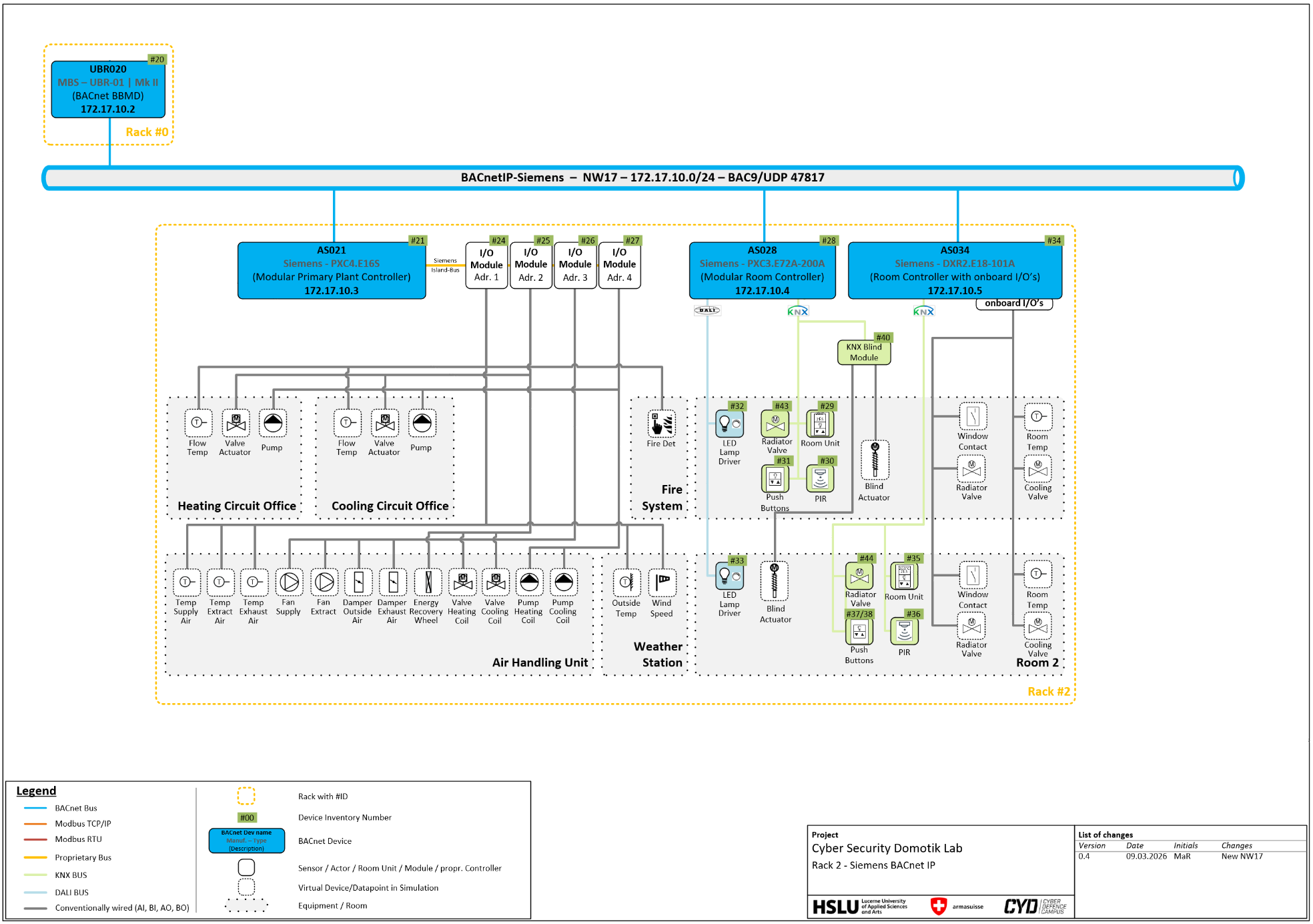}
\caption{Rack 2 architecture illustrating the relationship between Siemens BACnet/IP controllers, DALI lighting infrastructure, KNX-related components, sensors, and room-level automation devices.}
\label{fig:rack_arch}
\end{figure*}

\begin{figure}[t]
\centering
\includegraphics[width=\linewidth]{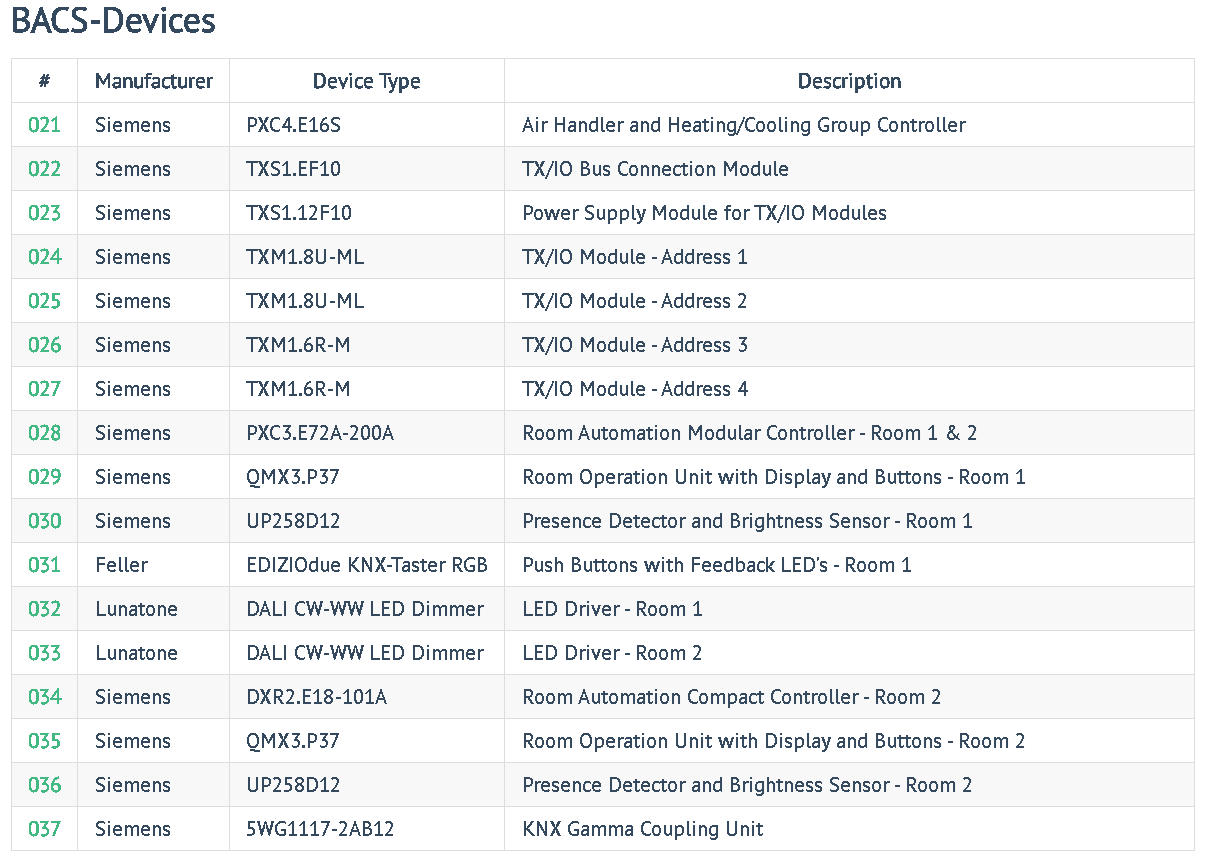}
\caption{Rack 2 device inventory. DALI devices 032 and 033 are documented as Lunatone DALI CW-WW LED dimmers for Room 1 and Room 2.}
\label{fig:devices}
\end{figure}

Figure~\ref{fig:overview} shows the provided overview of the building automation testbed. Figure~\ref{fig:rack_arch} focuses on Rack 2, the Siemens BACnet/IP rack used for the most relevant lighting and domotics analysis. Figure~\ref{fig:devices} shows the device inventory, including DALI devices associated with Room 1 and Room 2.

The relevant DALI lighting devices are summarized in Table~\ref{tab:dali_devices}.

\begin{table}[h]
\centering
\caption{Physical DALI lighting devices identified on Rack 2.}
\label{tab:dali_devices}
\begin{tabular}{llll}
\toprule
ID & Vendor & Device Type & Function \\
\midrule
032 & Lunatone & DALI CW-WW LED Dimmer & LED Driver -- Room 1 \\
033 & Lunatone & DALI CW-WW LED Dimmer & LED Driver -- Room 2 \\
\bottomrule
\end{tabular}
\end{table}

\section{BACnet Enumeration and Network-Level View}

\begin{figure}[t]
\centering
\includegraphics[width=\linewidth]{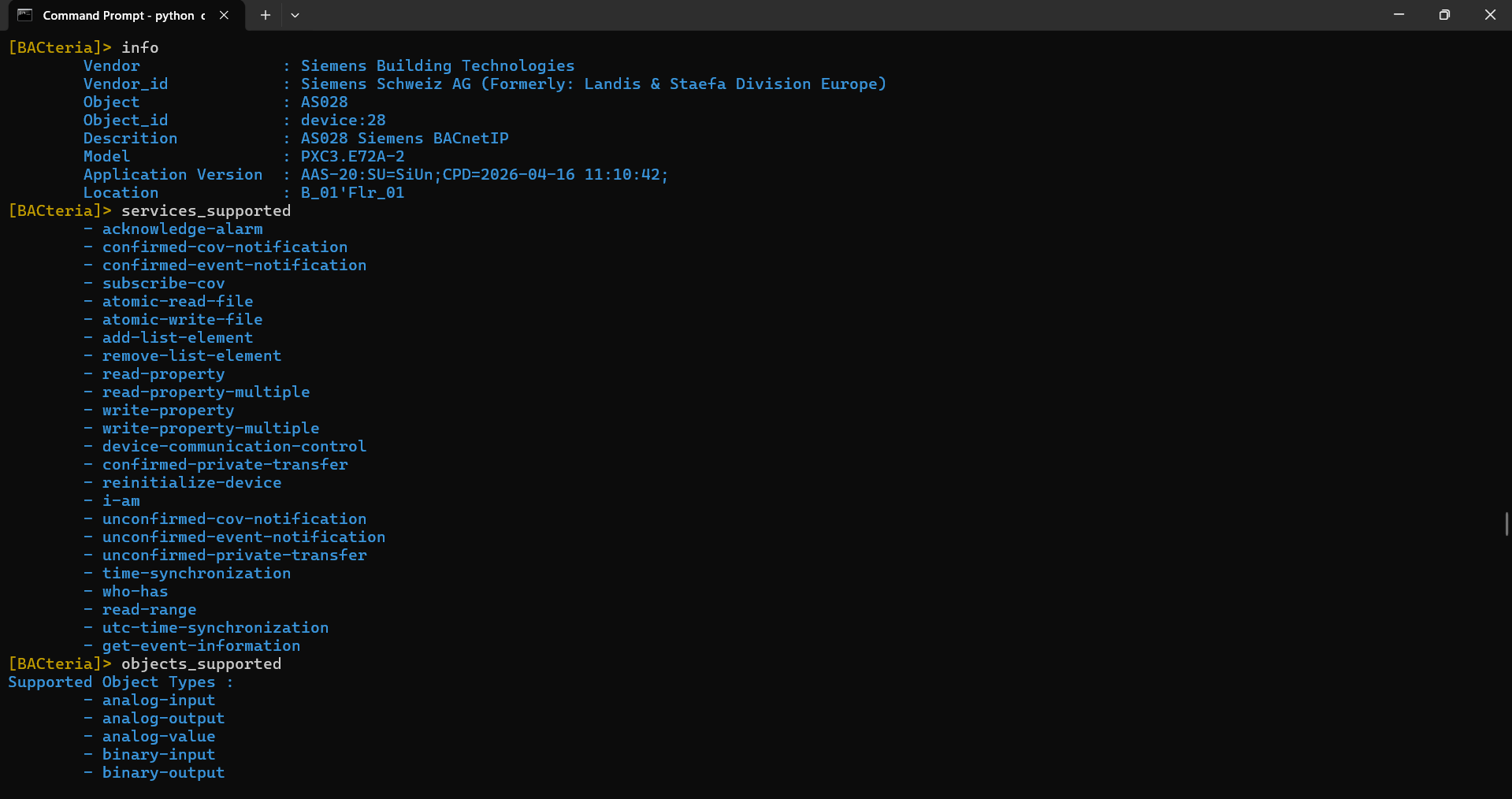}
\caption{BACteria output showing device fingerprinting, supported services, and supported object types.}
\label{fig:bacteria_info}
\end{figure}

The target controller was accessed using BACteria with a known BACnet/IP address and port. Device fingerprinting identified the target as a Siemens Building Technologies controller:

\begin{lstlisting}[style=terminal,caption={BACteria device fingerprint excerpt.}]
Vendor      : Siemens Building Technologies
Object      : AS028
Object_id   : device:28
Description : AS028 Siemens BACnetIP
Model       : PXC3.E72A-2
\end{lstlisting}

The supported BACnet services included read, write, event, file, time, and device-control operations:

\begin{lstlisting}[style=terminal,caption={Selected supported BACnet services.}]
read-property
read-property-multiple
write-property
write-property-multiple
device-communication-control
reinitialize-device
time-synchronization
utc-time-synchronization
get-event-information
\end{lstlisting}

Figure~\ref{fig:bacteria_info} illustrates the command-line enumeration workflow. The presence of \texttt{write-property}, \texttt{write-property-multiple}, \texttt{device-communication-control}, and \texttt{reinitialize-device} is relevant from a security-assessment perspective because these services represent control-relevant capabilities. However, service support alone does not imply unrestricted operational impact. Object permissions, priority arrays, automation logic, and physical system state must also be considered.

The device advertised object types including analog inputs, analog outputs, analog values, binary inputs, binary outputs, binary values, multi-state values, structured views, schedules, programs, trend logs, event logs, and notification classes. This diversity is technically useful but cognitively demanding: analysts must distinguish physical sensors from control points, status indicators from commands, and room abstractions from direct device objects.

\section{Threat Modeling Perspective}

This case study does not assign vulnerability scores or claim general vendor-wide weaknesses. Instead, it uses threat modeling to organize the observed attack surface. In a BACnet/IP testbed, potential risk areas include:

\begin{itemize}
    \item \textbf{Enumeration exposure}: device identity, model, object types, and room structures may be visible through BACnet property reads.
    \item \textbf{Control-service exposure}: support for \texttt{write-property} and related services indicates that commandable surfaces exist and require careful governance.
    \item \textbf{Object-level ambiguity}: not all objects are equally sensitive, and object names may not clearly reveal whether they represent sensors, states, groups, or commands.
    \item \textbf{Priority misunderstanding}: accepted writes may not affect physical devices if stronger priorities dominate.
    \item \textbf{Human error}: novice users may confuse object instances, physical labels, value types, and priority indexes.
\end{itemize}

This threat model supports a balanced interpretation: the environment exposes meaningful inspection and control semantics, but safe assessment requires understanding the protocol and the physical process.

\section{Object Hierarchy Reconstruction}

A key part of the analysis involved reconstructing the relationship between room-level structured views and lighting-control objects. The following command inspected a room-segment state object:

\begin{lstlisting}[style=terminal]
properties multi-state-value:33
\end{lstlisting}

The object was identified as:

\begin{lstlisting}[style=terminal,caption={Room segment state object.}]
object-identifier : multi-state-value:33
object-name       : B_01'Flr_01'RSegm2'RSegmSta
description       : Room segment state
present-value     : 1
state-text        : [Normal, Intervention active,
                     Fault, Alarm, Fault & alarm]
Node reference    : structured-view:76
\end{lstlisting}

The node reference led to \texttt{structured-view:76}, describing Room Segment 2:

\begin{lstlisting}[style=terminal,caption={Structured view for Room Segment 2.}]
object-identifier       : structured-view:76
object-name             : B_01'Flr_01'RSegm2
description             : Raumsegment Raum 2
subordinate-annotations : [RCtl, ROpUnCtl,
                           RSegmSta, RSegmStaMon,
                           Shd(2), Lgt(2)]
subordinate-list        : [...,
                           structured-view:170]
\end{lstlisting}

The subordinate annotation \texttt{Lgt(2)} pointed to \texttt{structured-view:170}, representing the Room 2 lighting branch:

\begin{lstlisting}[style=terminal,caption={Room 2 lighting structured view.}]
object-identifier       : structured-view:170
object-name             : B_01'Flr_01'RSegm2'Lgt(2)
description             : Lighting 2
subordinate-annotations : [LgtCmd(2), ROpMod,
                           LocManOp, PscOp,
                           GrnLf, CenEmgLgt]
subordinate-list        : [260:8,
                           structured-view:171,
                           structured-view:173,
                           structured-view:178,
                           ...]
\end{lstlisting}

\begin{figure}[t]
\centering
\includegraphics[width=\linewidth]{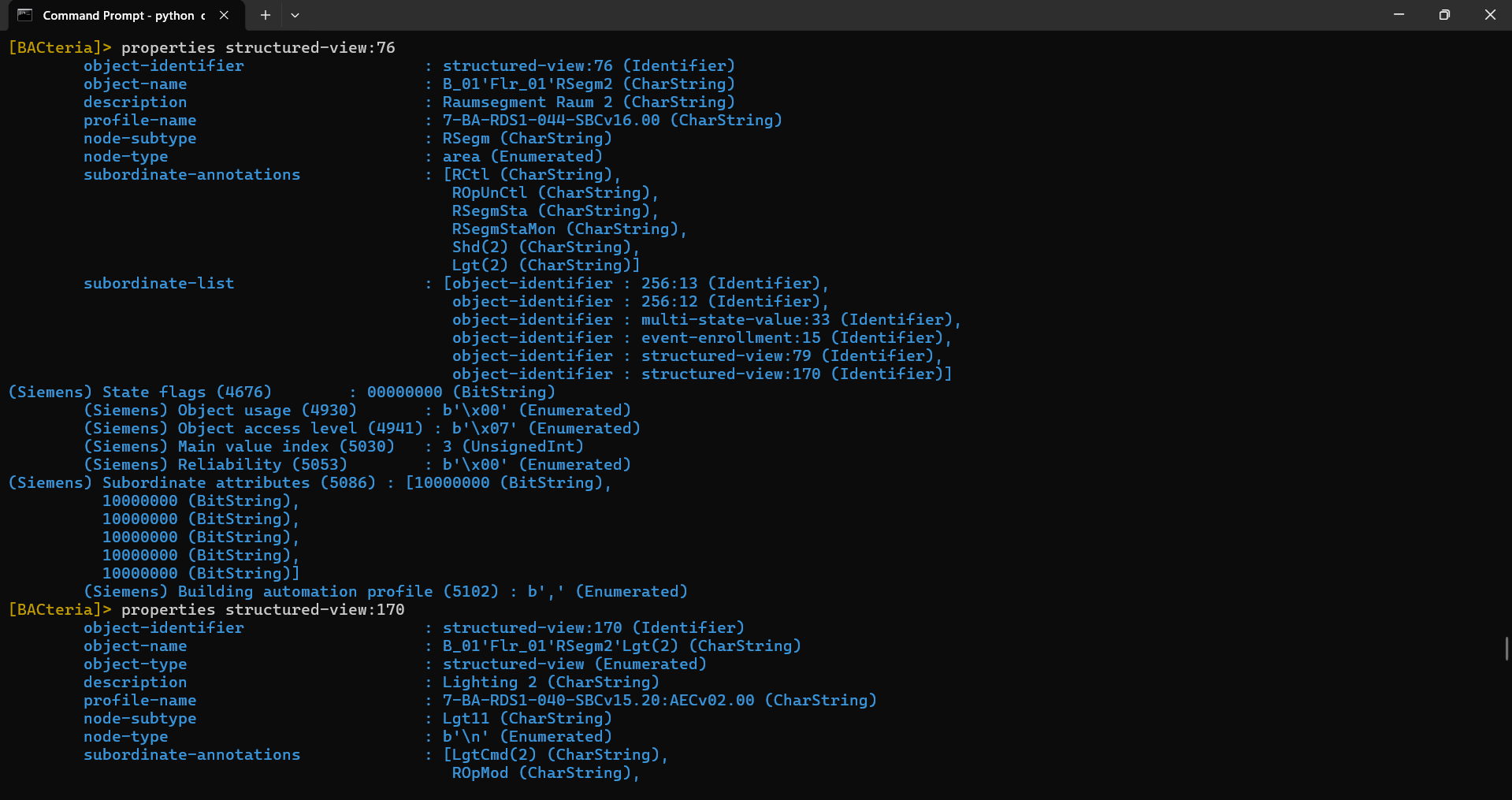}
\caption{BACteria inspection of the Room 2 lighting structured view. The subordinate list includes \texttt{260:8}, the lighting command object.}
\label{fig:hierarchy}
\end{figure}

This traversal shows how a room state object can lead to an area node, then to a lighting node, and finally to command-related subordinate objects. Figure~\ref{fig:hierarchy} shows the Room 2 structured-view analysis.

\section{BACnet-to-DALI Control Mapping}

\begin{figure}[t]
\centering
\includegraphics[width=\linewidth]{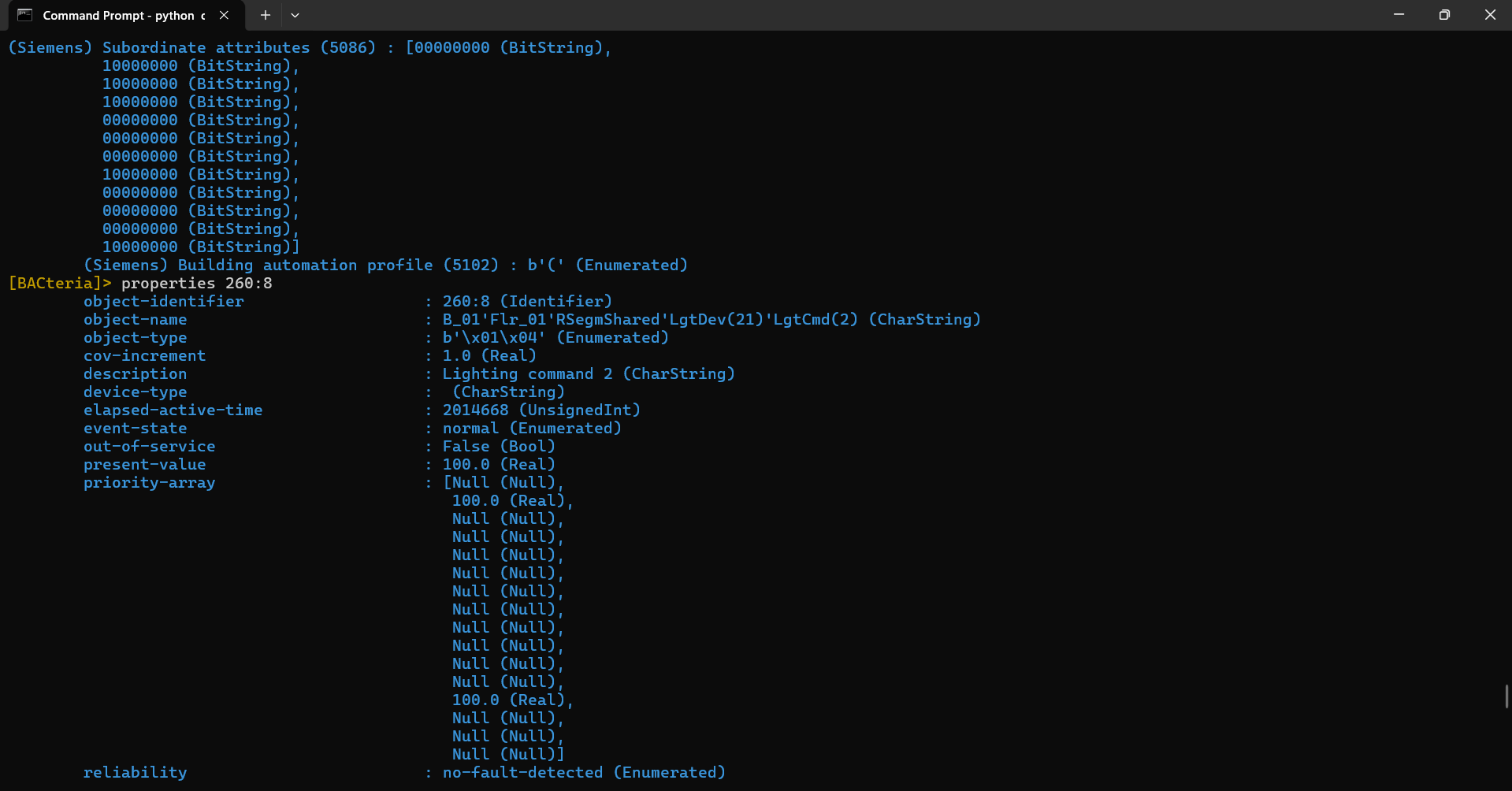}
\caption{BACnet object \texttt{260:8}, identified as \texttt{Lighting command 2}, with I/O address \texttt{DALI-1:G=2}.}
\label{fig:dali_command}
\end{figure}

The most relevant object in the Room 2 lighting branch was \texttt{260:8}. Inspecting it produced the following:

\begin{lstlisting}[style=terminal,caption={Lighting command object linked to DALI Group 2.}]
object-identifier : 260:8
description       : Lighting command 2
present-value     : 100.0
I/O address       : DALI-1:G=2
Present priority  : 2
Tracking value    : 100.0
\end{lstlisting}

This object connected the BACnet object hierarchy to DALI group-level infrastructure. The physical rack inventory identified DALI devices 032 and 033 as Lunatone DALI CW-WW LED dimmers for Room 1 and Room 2. The documented protocol-level interpretation therefore follows the chain:

\begin{lstlisting}[style=terminal]
Room Segment 2
  -> Lighting 2
     -> Lighting command 2
        -> DALI-1:G=2
           -> Room 2 DALI lighting infrastructure
\end{lstlisting}

Figure~\ref{fig:dali_command} shows the inspected lighting command object.

\section{Controlled Write Observation and Priority Semantics}

A controlled property write was performed in the authorized testbed using BACteria:

\begin{lstlisting}[style=terminal]
set_property 260:8 present-value Real 50 8 8
Property was updated
\end{lstlisting}

This observation must be interpreted carefully. The inspected object also reported \texttt{Present priority: 2}, while the command used priority 8. In BACnet semantics, lower priority numbers dominate higher priority numbers. Therefore, a priority-8 write can be processed without becoming the effective physical command if a higher-priority value is active.

This distinction separates three layers:

\begin{enumerate}
    \item \textbf{Service availability}: the device supports write-property operations.
    \item \textbf{Object-level processing}: the target object can process a write request.
    \item \textbf{Effective control}: the physical outcome depends on the active priority and automation logic.
\end{enumerate}

\begin{figure}[t]
\centering
\includegraphics[width=\linewidth]{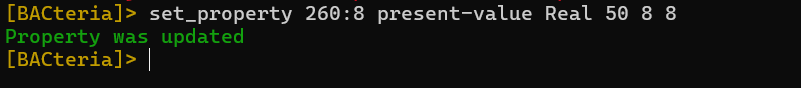}
\caption{BACteria confirmation of a controlled property write to the lighting command object.}
\label{fig:write}
\end{figure}

\begin{figure}[t]
\centering
\includegraphics[width=\linewidth]{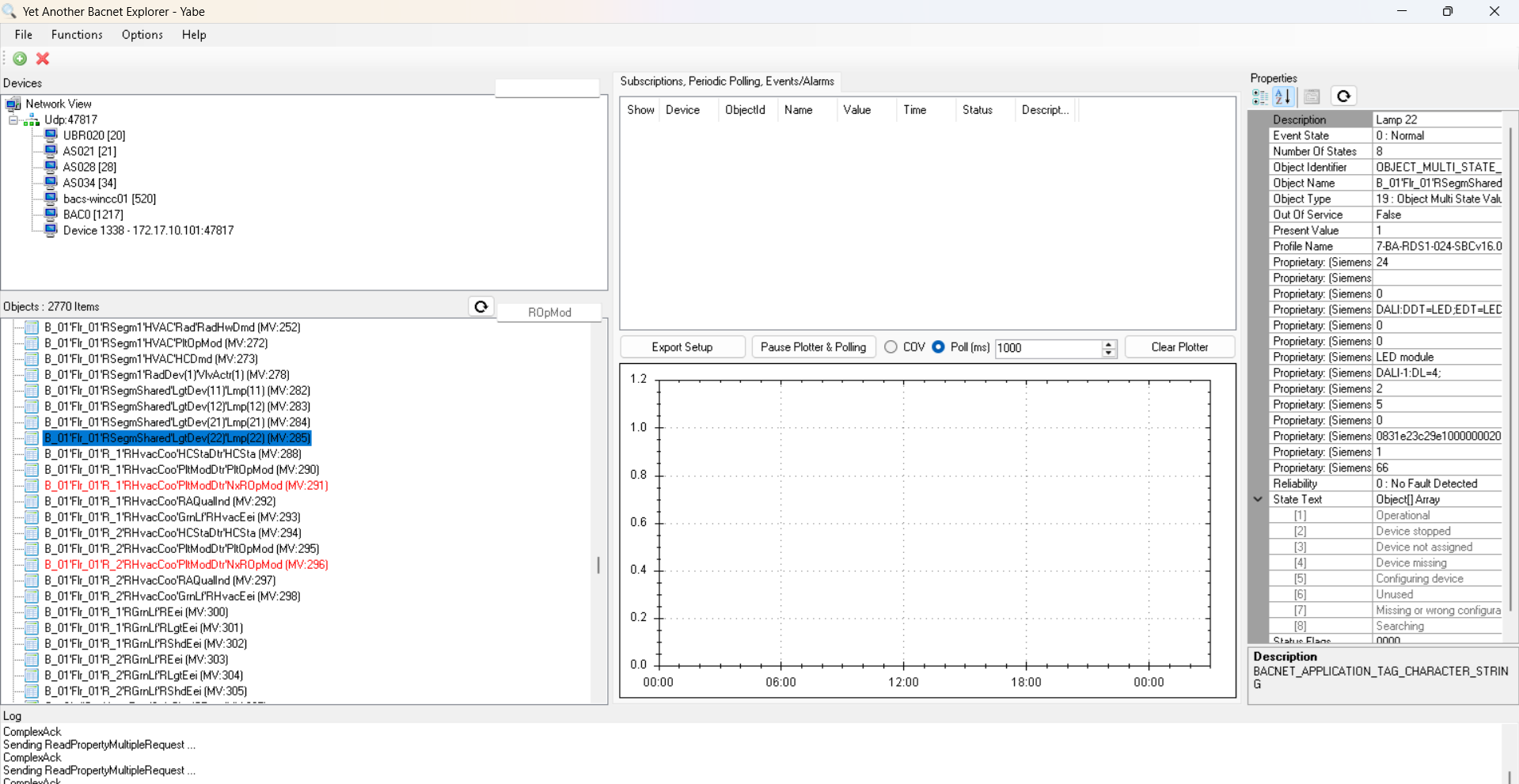}
\caption{Yabe BACnet explorer view used for graphical inspection of objects and properties.}
\label{fig:yabe}
\end{figure}

Figure~\ref{fig:write} shows the terminal confirmation. Figure~\ref{fig:yabe} shows Yabe being used as a graphical companion for object inspection.

\section{Human-Centered Analysis}

\subsection{Tool Learnability}

BACnet tools expose low-level protocol detail that is powerful but initially difficult for inexperienced users. A participant must interpret object identifiers, object types, property names, value types, priorities, and physical labels. In this case, progress depended on switching between Yabe's visual browser, BACteria's command-line output, rack diagrams, and physical device labels.

This indicates that BACS security tools could benefit from stronger role-based visualization. For example, a tool could visually distinguish sensors, commandable values, group controls, structured views, and physical devices.

\subsection{Physical Observability}

\begin{figure}[t]
\centering
\begin{subfigure}{0.48\linewidth}
\centering
\includegraphics[width=\linewidth]{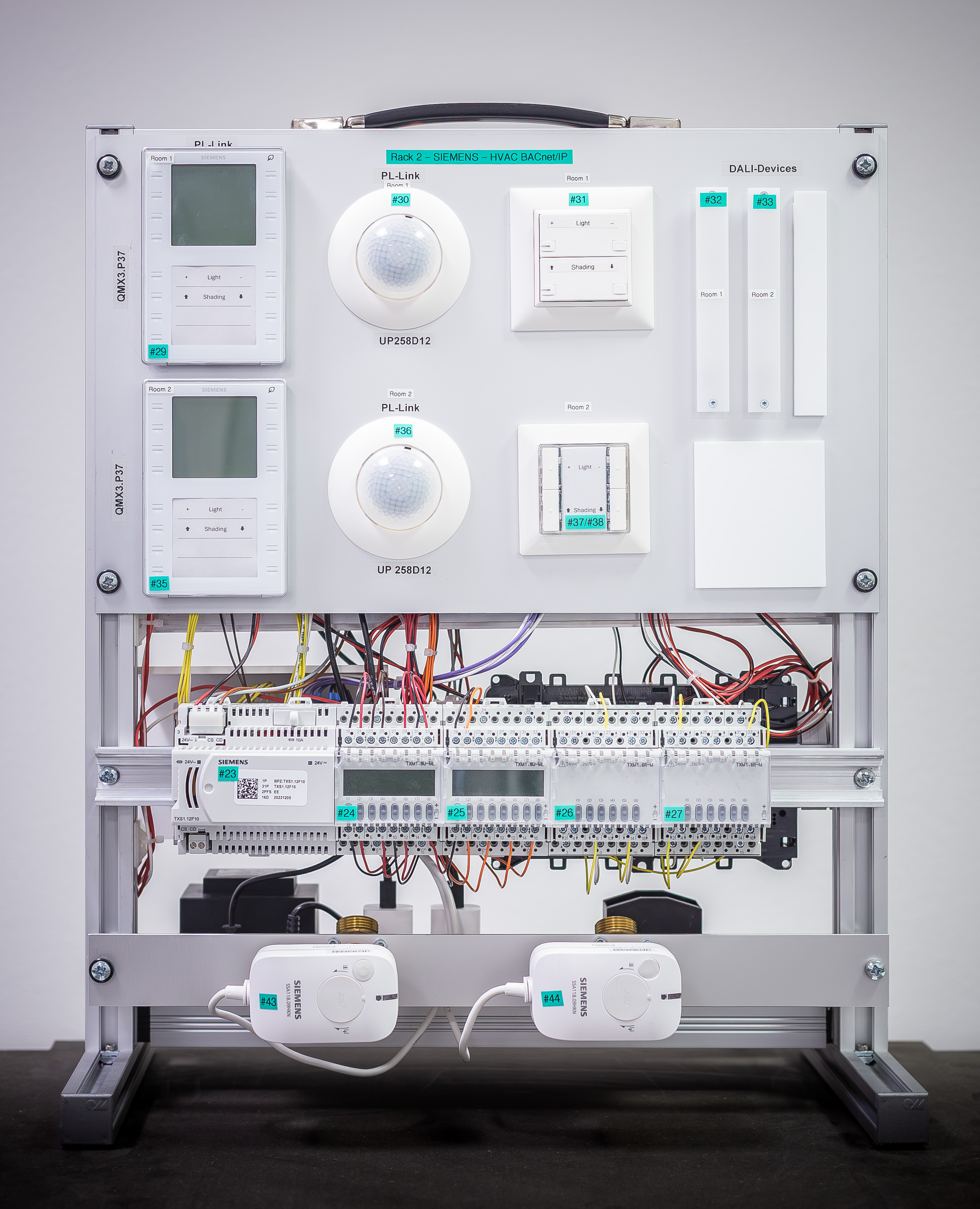}
\caption{Front view}
\end{subfigure}
\hfill
\begin{subfigure}{0.48\linewidth}
\centering
\includegraphics[width=\linewidth]{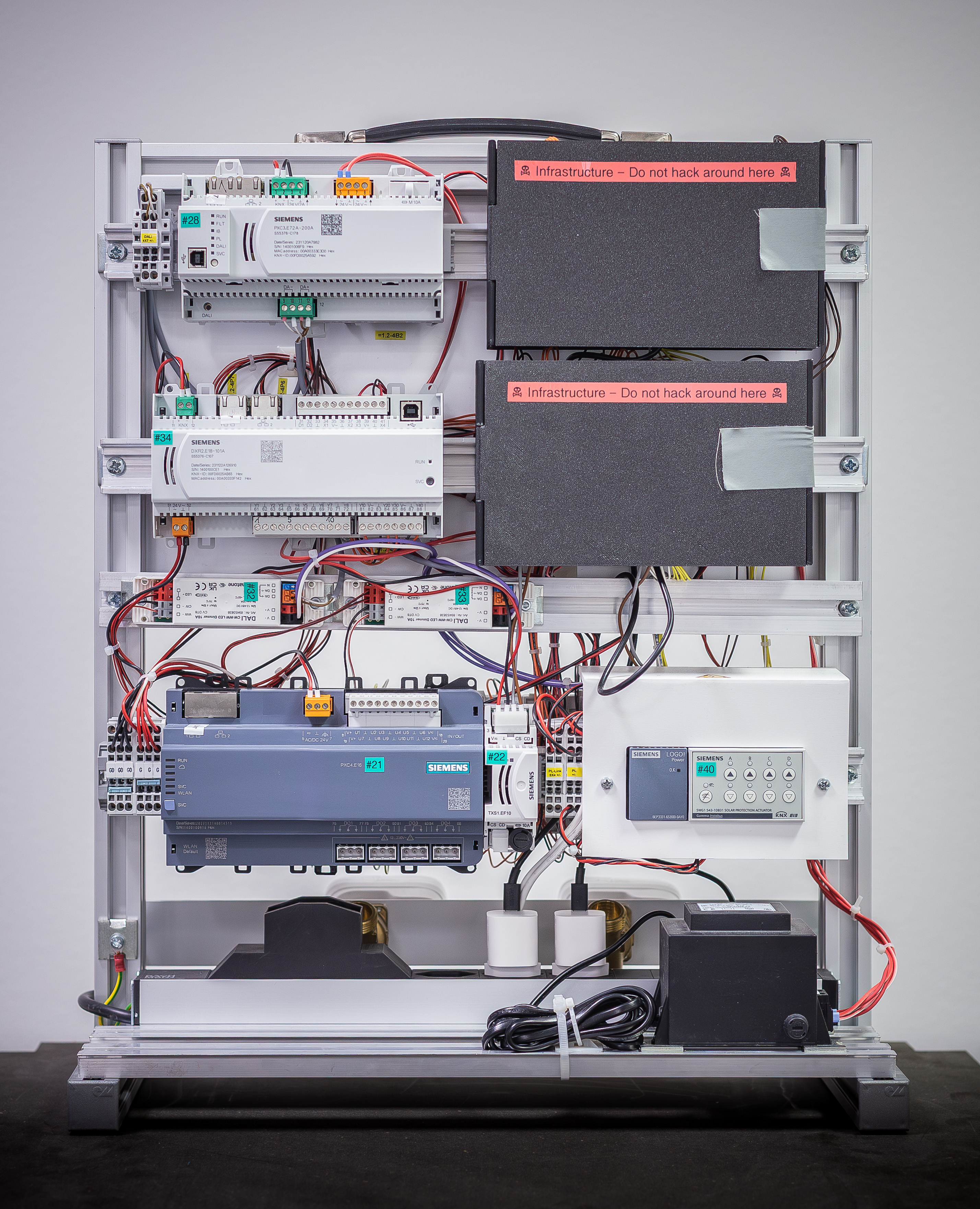}
\caption{Rear view}
\end{subfigure}
\caption{Physical Rack 2 setup used to correlate protocol-level analysis with observable hardware.}
\label{fig:rack_photos}
\end{figure}

Cyber-physical systems differ from ordinary software systems because digital commands may or may not produce immediately visible effects. A lighting value may be overridden by automation priority, mapped to a DALI group instead of a single lamp, or visible only on a specific rack. The rack photographs in Figure~\ref{fig:rack_photos} show why physical context matters for interpretation.

\subsection{Cognitive Load and Safety}

The workflow required reasoning about network addressing, BACnet object notation, structured-view hierarchies, DALI group identifiers, command priorities, and physical devices. This creates substantial cognitive load. For educational BACS environments, safety-oriented interface features could include warnings for high-priority control points, explanations of \texttt{priority-array} semantics, safer read-only exploration modes, and guided mapping from protocol objects to rack labels.

\section{Lessons Learned}

Several lessons emerged from the case study:

\begin{itemize}
    \item \textbf{Structured views are valuable navigation aids.} Traversing \texttt{structured-view} objects provided a systematic path from room-level abstractions to lighting-specific control objects.
    \item \textbf{Physical labels and protocol identifiers do not always align.} Rack labels such as 032 and 033 helped identify physical DALI devices, but BACnet control paths used different object identifiers.
    \item \textbf{Priority arrays prevent simplistic interpretations.} A processed write does not automatically imply physical effect when higher-priority values dominate.
    \item \textbf{Combining tools improves understanding.} Yabe supported visual exploration, while BACteria supported precise command-line enumeration.
    \item \textbf{Cybersecurity education in BACS requires HCI support.} Learners need interfaces that explain risk, role, and control semantics, not only raw protocol fields.
\end{itemize}

Beyond the individual observations, the case study suggests that BACS security education benefits from workflows that explicitly connect protocol-level evidence with physical-system interpretation. In particular, learners need support for moving between object identifiers, commandable properties, automation priorities, physical labels, and observable rack behavior. 

\section{Discussion}

The case study illustrates that BACS security assessment requires both technical and human-centered reasoning. Technically, the path from \texttt{multi-state-value:33} to \texttt{structured-view:76}, then to \texttt{structured-view:170}, and finally to \texttt{260:8} demonstrates how BACnet object hierarchies can expose meaningful relationships between room abstractions, lighting functions, and DALI group-level infrastructure.

At the same time, the HCI dimension is essential. Without physical diagrams, rack labels, and careful tool interpretation, the relationship between \texttt{DALI-1:G=2} and Room 2 lighting infrastructure would be difficult to understand. The participant's workflow was therefore not just network enumeration, but sensemaking across digital and physical layers.

The priority behavior further shows why careful wording is required in academic reporting. A write confirmation indicates that the request was processed. It does not alone establish full physical control. In building automation, operational meaning depends on object permissions, priority arrays, supervisory logic, and the physical process. From a design perspective, this suggests that future BACS assessment tools should support layered explanations rather than only exposing raw protocol fields.

\section{Ethical Considerations and Limitations}

All observations described in this paper were conducted in an authorized educational hackathon testbed. The analysis was limited to the provided environment and focused on documentation, enumeration, interpretation, and controlled interaction. No claims are made about unauthorized third-party systems. No destructive modification, persistence, or out-of-scope targeting is included. The purpose is to support cybersecurity education, responsible experimentation, and safer design of BACS tools and training environments.

The paper intentionally avoids presenting the work as an exploitation guide. Commands, object identifiers, and screenshots are included only where they support the reconstruction of BACnet object hierarchies, DALI mapping, and priority semantics inside the authorized testbed. Sensitive operational conclusions are therefore framed at the level of system interpretation, educational value, and responsible assessment rather than generalizable vulnerability disclosure. This framing is important because building automation systems are cyber-physical environments in which careless experimentation may affect physical devices, safety assumptions, or shared training infrastructure.

This paper reports a single case study from an educational testbed. The observations should not be generalized to all Siemens, BACnet, or DALI deployments. The analysis is based on available screenshots, terminal outputs, physical inspection, and participant notes. It does not provide statistical evaluation, large-scale scanning, vendor-wide security assessment, or complete BACnet/SC security evaluation.

\section{Conclusion}

This paper presented a cybersecurity and human-centered case study of BACnet-controlled DALI infrastructure in an educational building automation testbed. The analysis documented BACnet service exposure, supported object types, structured-view traversal, DALI group-level mapping, and BACnet priority semantics. It also showed that tool usability, physical observability, and cognitive load strongly influence how participants interpret and safely interact with cyber-physical building systems.

The central contribution is the reconstruction of the relationship between BACnet services, structured room objects, DALI group addressing, and physical lighting infrastructure. Rather than treating the testbed only as a network target, the paper frames BACS assessment as a combined process of protocol analysis, physical-system interpretation, and human-centered sensemaking. This combined cyber-physical and HCI perspective is important for BACS security education, responsible experimentation, and the design of safer analysis tools.

\balance
\bibliographystyle{IEEEtran}
\bibliography{references}

@misc{bacnet_committee,
  author       = {{BACnet Committee}},
  title        = {{BACnet Committee -- ASHRAE SSPC 135}},
  year         = {{n.d.}},
  howpublished = {\url{https://bacnet.org/}},
  note         = {Accessed: 2026-06-02}
}

@misc{ashrae_bacnet,
  author       = {{ASHRAE}},
  title        = {{BACnet: The ASHRAE Building Automation and Control Networking Protocol}},
  year         = {{n.d.}},
  howpublished = {\url{https://www.ashrae.org/technical-resources/bookstore/bacnet}},
  note         = {Accessed: 2026-06-02}
}

@misc{dali_alliance,
  author       = {{DALI Alliance}},
  title        = {{DALI and DALI-2: Standardized Smart Lighting Control and IEC 62386}},
  year         = {{n.d.}},
  howpublished = {\url{https://www.dali-alliance.org/dali/}},
  note         = {Accessed: 2026-06-02}
}

@article{graveto2022bacs,
  author  = {Vitor Graveto and Tiago Cruz and Paulo Sim{\~o}es},
  title   = {Security of Building Automation and Control Systems: Survey and Future Research Directions},
  journal = {Computers \& Security},
  volume  = {112},
  pages   = {102527},
  year    = {2022},
  doi     = {10.1016/j.cose.2021.102527},
  note    = {\url{https://doi.org/10.1016/j.cose.2021.102527}}
}

@article{morales2024bas,
  author  = {Christopher Morales-Gonzalez and Matthew Harper and Michael Cash and Lan Luo and Zhen Ling and Qun Z. Sun and Xinwen Fu},
  title   = {On Building Automation System Security},
  journal = {High-Confidence Computing},
  volume  = {4},
  number  = {3},
  pages   = {100236},
  year    = {2024},
  doi     = {10.1016/j.hcc.2024.100236},
  note    = {\url{https://doi.org/10.1016/j.hcc.2024.100236}}
}

@misc{armasuisse_bacs_2025,
  author       = {{armasuisse Science and Technology}},
  title        = {{CYD Campus BACS Hackathon 2025 -- Exploring Security for Building Automation and Control Systems}},
  year         = {2025},
  howpublished = {\url{https://www.ar.admin.ch/en/domotic-hackathon-cyd-campus-en}},
  note         = {Accessed: 2026-06-02}
}

@article{li2023critical,
  author  = {Guowen Li and Lingyu Ren and Yangyang Fu and Zhiyao Yang and Veronica Adetola and Jin Wen and Qi Zhu and Teresa Wu and K. Selcuk Candan and Zheng O'Neill},
  title   = {A Critical Review of Cyber-Physical Security for Building Automation Systems},
  journal = {Annual Reviews in Control},
  volume  = {55},
  pages   = {237--254},
  year    = {2023},
  doi     = {10.1016/j.arcontrol.2023.02.004},
  note    = {\url{https://doi.org/10.1016/j.arcontrol.2023.02.004}}
}

@inproceedings{gasser2017bacnet,
  author    = {Oliver Gasser and Quirin Scheitle and Carl Denis and Nadja Schricker and Georg Carle},
  title     = {Security Implications of Publicly Reachable Building Automation Systems},
  booktitle = {2017 IEEE Security and Privacy Workshops (SPW)},
  pages     = {199--204},
  address   = {San Jose, CA, USA},
  year      = {2017},
  doi       = {10.1109/SPW.2017.13},
  note      = {\url{https://doi.org/10.1109/SPW.2017.13}}
}

@misc{bacnet_sc_whitepaper,
  author       = {{ASHRAE}},
  title        = {{BACnet Secure Connect Whitepaper}},
  year         = {2019},
  howpublished = {\url{https://www.ashrae.org/File%20Library/Technical%20Resources/Bookstore/BACnet-SC-Whitepaper-v15_Final_20190521.pdf}},
  note         = {Accessed: 2026-06-02}
}

@article{grobler2021human,
  author  = {Marthie Grobler and Raj Gaire and Surya Nepal},
  title   = {User, Usage and Usability: Redefining Human Centric Cyber Security},
  journal = {Frontiers in Big Data},
  volume  = {4},
  year    = {2021},
  doi     = {10.3389/fdata.2021.583723},
  note    = {\url{https://doi.org/10.3389/fdata.2021.583723}}
}

@article{yamin2020cyberranges,
  author  = {Muhammad Mudassar Yamin and Basel Katt and Vasileios Gkioulos},
  title   = {Cyber Ranges and Security Testbeds: Scenarios, Functions, Tools and Architecture},
  journal = {Computers \& Security},
  volume  = {88},
  pages   = {101636},
  year    = {2020},
  doi     = {10.1016/j.cose.2019.101636},
  note    = {\url{https://doi.org/10.1016/j.cose.2019.101636}}
}

\end{document}